\documentclass[aps,reprint,superscriptaddress,floatfix]{revtex4-1}

\usepackage[urlcolor=blue,colorlinks=true,citecolor=blue,linkcolor=blue,pdfstartview={FitH},bookmarks=false]{hyperref}

\usepackage[utf8]{inputenc}
\usepackage[T1]{fontenc}
\usepackage{graphicx}
\begin{document}

\title{\textit {Ab initio} and nuclear inelastic scattering studies of Fe$_3$Si/GaAs heterostructures}

\author{O.~Sikora}
\email[e-mail: ]{olga.sikora@ifj.edu.pl}
\affiliation{Institute of Nuclear Physics, Polish Academy of Sciences, PL-31342 Kraków, Poland}

\author{J.~Kalt}
\affiliation{Laboratory for Applications of Synchrotron Radiation, Karlsruhe Institute of Technology, D-76131 Karlsruhe, Germany}
\affiliation{Institute for Photon Science and Synchrotron Radiation, Karlsruhe Institute of Technology, D-76344 Eggenstein-Leopoldshafen, Germany}

\author{M.~Sternik}
\affiliation{Institute of Nuclear Physics, Polish Academy of Sciences, PL-31342 Kraków, Poland}
\author{A.~Ptok}
\affiliation{Institute of Nuclear Physics, Polish Academy of Sciences, PL-31342 Kraków, Poland}
\author{P.~T.~Jochym}
\affiliation{Institute of Nuclear Physics, Polish Academy of Sciences, PL-31342 Kraków, Poland}
\author{J.~\L{}a\.{z}ewski}
\affiliation{Institute of Nuclear Physics, Polish Academy of Sciences, PL-31342 Kraków, Poland}
\author{K.~Parlinski}
\affiliation{Institute of Nuclear Physics, Polish Academy of Sciences, PL-31342 Kraków, Poland}
\author{P.~Piekarz}
\affiliation{Institute of Nuclear Physics, Polish Academy of Sciences, PL-31342 Kraków, Poland}
\author{I.~Sergueev}
\affiliation{Deutsches Elektronen-Synchrotron, D-22607 Hamburg, Germany}
\author{H.-C.~Wille}
\affiliation{Deutsches Elektronen-Synchrotron, D-22607 Hamburg, Germany}
\author{J.~Herfort}
\affiliation{Paul-Drude-Institut für Festkörperelektronik, D-10117 Berlin, Germany}
\author{B.~Jenichen}
\affiliation{Paul-Drude-Institut für Festkörperelektronik, D-10117 Berlin, Germany}
\author{T.~Baumbach}
\affiliation{Laboratory for Applications of Synchrotron Radiation, Karlsruhe Institute of Technology, D-76131 Karlsruhe, Germany}
\affiliation{Institute for Photon Science and Synchrotron Radiation, Karlsruhe Institute of Technology, D-76344 Eggenstein-Leopoldshafen, Germany}
\author{S.~Stankov}
\affiliation{Laboratory for Applications of Synchrotron Radiation, Karlsruhe Institute of Technology, D-76131 Karlsruhe, Germany}
\affiliation{Institute for Photon Science and Synchrotron Radiation, Karlsruhe Institute of Technology, D-76344 Eggenstein-Leopoldshafen, Germany}

\date{\today}

\begin{abstract}
The structure and dynamical properties of the Fe$_3$Si/GaAs(001) interface are investigated by density functional theory and nuclear inelastic scattering measurements.
The stability of four different atomic configurations of the Fe$_3$Si/GaAs multilayers is analyzed by calculating the formation energies and phonon dispersion curves. The differences in charge density, magnetization, and electronic density of states between the configurations are examined. 
Our calculations unveil that magnetic moments of the Fe atoms tend to align in a plane parallel to the interface, along the [110] direction of the Fe$_3$Si crystallographic unit cell. 
In some configurations, the spin polarization of interface layers is larger than that of bulk Fe$_3$Si. 
The effect of the interface on element-specific and layer-resolved phonon density of states is discussed.
The Fe-partial phonon density of states measured for the Fe$_3$Si layer thickness of three monolayers is compared with theoretical results obtained for each interface atomic configuration. 
The best agreement is found for one of the configurations with a mixed Fe-Si interface layer, which reproduces the anomalous enhancement of the phonon density of states below 10 meV.

\end{abstract}

\maketitle

\section{Introduction}

The progress in magnetoelectric and spintronic applications depends on the development of high-quality hybrid systems connecting ferromagnetic (FM) and semiconducting (SC) materials.
In such systems, the FM electrode provides a spin-polarized current, which can be injected into the SC layer~\cite{Prinz1995,Zhu2001,Hanbicki2002,Ramsteiner2002}.
Highly efficient spin-transport can be achieved only in FM/SC interfaces or multilayers with particular crystalline properties~\cite{Ploog2002,Dash2011}.
The presence of the interface can strongly modify thermo-elastic properties of the contact area and determine the stability conditions of the entire system. 
It may also alter the coupling between electrons and atomic vibrations (phonons) and thus influence
the spin transport (e.g. spin-flip processes)~\cite{Pradip2016}. Therefore, a full characterization of the interface properties, 
including the lattice dynamics, becomes a precondition for further developments in this area. 

Phonon properties of nanomaterials can strongly differ from those of bulk crystals \cite{Bohnen1993,Fultz1997,Kara1998,Roldan2007,Stankov2008,Couet2010}. 
Forces acting between atoms located at the interface or surface are modified due to changes in local bonding geometry or interatomic distances (epitaxial strain) \cite{Stankov2007,Slezak2007}.    
Other effects related to finite sizes and reduced dimensionality also influence atomic vibrations \cite{Spiridis2015}.
These modifications of lattice dynamics in nanosystems lead to new phenomena such as phonon folding~\cite{Colvard1980}, phonon localization and confinement~\cite{Roldan2008,Seiler2016,Keune2018} as well as coherent and enhanced heat conductivity~\cite{Luckyanova2012,Schroeder2015,Park2017}.
Further studies on these effects may pave the way to phonon engineering~\cite{Balandin2012,Volz2016} and efficient thermal management in nanoscale devices~\cite{Pop2010,Moore2014}.

Fe$_3$Si is a ferromagnetic material with a high Curie temperature and a Heusler alloy structure~\cite{Hines1976}.
Due to almost perfect matching of lattice parameters, Fe$_3$Si deposited epitaxially on a GaAs(001) surface 
shows high interfacial quality and high thermal stability up to 700~K~\cite{Herfort2003,Herfort2004,Herfort2005}. 
These properties together with stable magnetization close to the bulk value~\cite{Ionescu2005,Lenz2005,Lenz2006,Herfort2008} 
make this system a suitable candidate for spintronic applications.   
Indeed, spin injection from a Fe$_3$Si layer was observed by the circular
polarization of the light emitted from a GaAs substrate~\cite{Kawaharazuka2004}.
Also, point contact Andreev reflection measurements show $45\pm5$\%~polarization of the transport current~\cite{Ionescu2005}. 

The properties of crystalline Fe$_3$Si  were studied previously using density functional theory (DFT)~\cite{Dennler2006,Hafner2007,Liang2011,Odkhuu2011,Sandalov2015}.
The calculated magnetic moments~\cite{Dennler2006,Odkhuu2011} and phonon dispersion relations~\cite{Dennler2006,Liang2011} showed good agreement with the experimental data~\cite{Hines1976,Randl1995}.
The studies performed for the Fe$_3$Si/GaAs(110) multilayer showed an increase of magnetic moments~\cite{Herper2008} and oscillation of the spin polarization with the interface distance~\cite{Grunebohm2010}.
M\"{o}ssbauer spectroscopy and DFT studies yielded evidence of some disorder or atomic interdiffusion at the Fe$_3$Si/GaAs(001) interface~\cite{Weis2010}. 
In our previous work we presented a theoretical and nuclear inelastic scattering (NIS) study on phonon properties of the Fe$_3$Si/GaAs(001) interface~\cite{Kalt2018}. 
In thin Fe$_3$Si layers we found a significant enhancement of the low-energy phonon states, in comparison to the bulk material. 
The DFT calculations explained the observed effect by the existence of interface-specific phonon states originating from the significantly reduced atomic force constants.

The main goal of this paper is to present systematic DFT studies of the structural, electronic, magnetic, and phonon properties of the Fe$_3$Si/GaAs(001) heterostructures with different atomic configurations, in order to predict the properties
of the Fe$_3$Si thin film on a GaAs substrate. We analyze the charge density and magnetic moments obtained for each layer of the Fe$_3$Si/GaAs system. We present also the layer-resolved electronic density of states and analyze the spin polarization at the Fermi level.
Based on the calculated phonon dispersion relations, we determine and discuss the dynamical stability of the interface depending on the atomic configurations.
For each configuration, we calculate the Fe- and Si-partial phonon density of states (PDOS) for in-plane and out-of-plane vibrations at the interface.
In the experiment, the Fe-partial PDOS of the thin Fe$_3$Si layer deposited on the GaAs(001) substrate was measured by NIS and compared to a theoretical model using PDOS calculated for the bulk Fe$_3$Si crystal and Fe$_3$Si/GaAs heterostructures with different atomic configurations.

The paper is organized as follows. In Secs.~\ref{sec:structural} and~\ref{sec:computational}, we present the theoretical models of the Fe$_3$Si/GaAs interface and calculation details, respectively. 
In Sec.~\ref{sec:nis}, the sample and experimental method are described.
Structural properties of the optimized models are discussed in Sec.~\ref{sec:theo_structural}. The magnetic moments and magnetization maps are presented in Sec.~\ref{sec:theo_magnetic}. In Secs.~\ref{sec:theo_charge} and~\ref{sec:theo_electron}, the results on charge distribution and electronic density of states are analyzed. 
Phonon properties of the interface are discussed in Sec.~\ref{sec:theo_dynamic}. 
The experimental results are presented and compared with the theory in Sec.~\ref{sec:experimental}.
The summary and conclusions are included in Sec.~\ref{sec:summary}. 

\section{Methodology}

\subsection{Structural models}
\label{sec:structural}

\begin{figure}[!t]
\centering
\includegraphics[width=\linewidth]{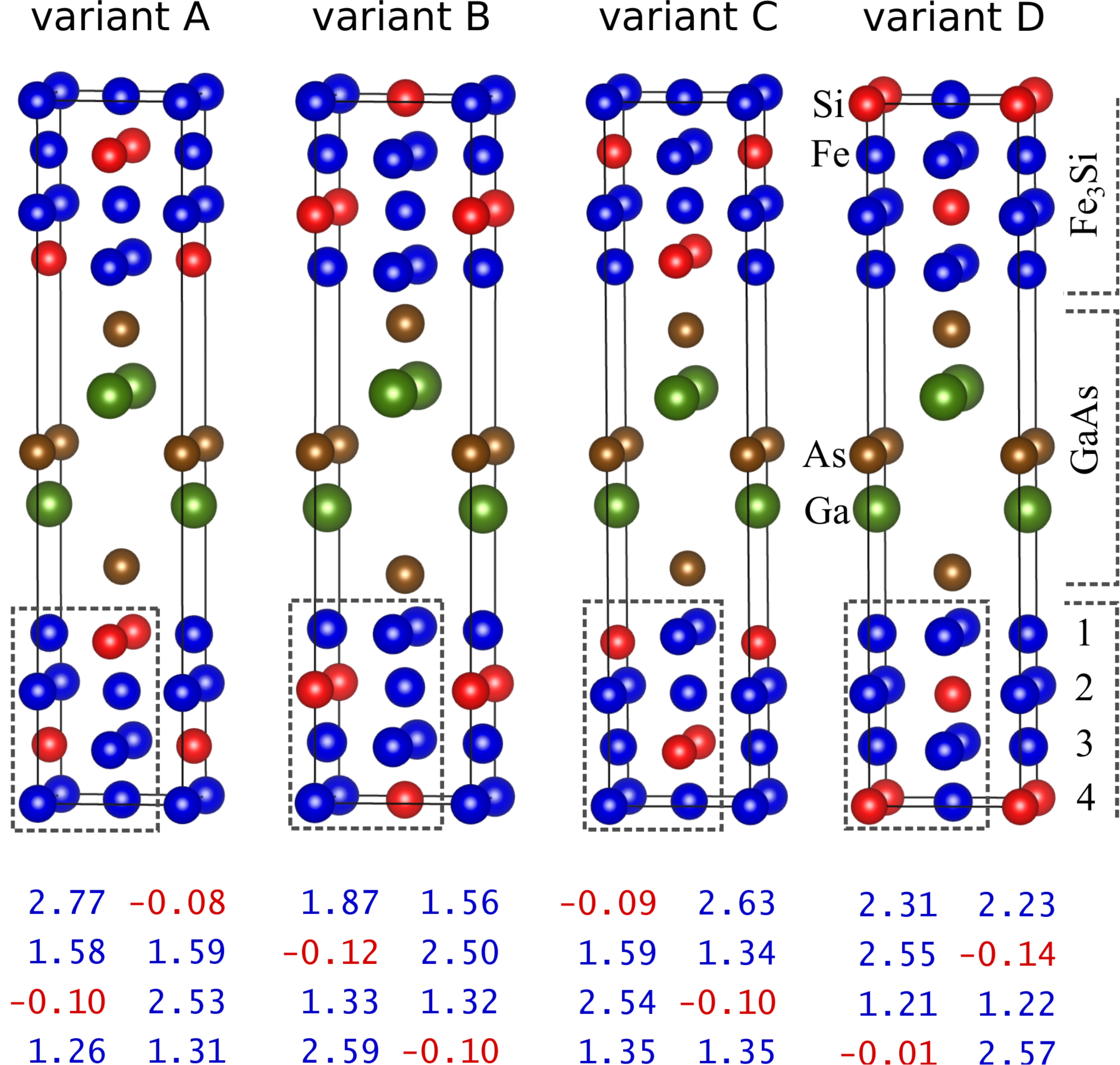}
\caption{\label{fig:structure} The primitive unit cells of four Fe$_3$Si/GaAs heterostructure variants A, B, C, and D. Blue, red, green, and brown balls depict iron, silicon, gallium, and arsenic atoms, respectively. The numbers on the right-hand side enumerate layers in the Fe$_3$Si block. The values at the bottom indicate magnetic moments of nonequivalent Fe and Si atoms inside the areas marked by dashed lines.
The image was rendered using {\sc VESTA} software \cite{vesta}.}
\label{fig:cells}
\end{figure}

Bulk Fe$_3$Si and GaAs crystals show the cubic $Fm\bar{3}m$ and $F\bar{4}3m$ space groups, respectively. The similar layered arrangement of atoms and almost equal lattice constants facilitate the formation of Fe$_3$Si/GaAs heteroepitaxial structures~\cite{Herfort2003,Herfort2004,Herfort2005,Jenichen2005}.
As shown previously~\cite{Kaganer2008}, there are four high-symmetry variants of the Fe$_3$Si/GaAs system. 
In the present paper, the structures based on these variants are used as models of the Fe$_3$Si/GaAs interfaces. 
In Fig.~\ref{fig:structure}, the primitive cells of the considered systems are presented.  
Each cell contains 12 layers and can be divided into a block of 5 layers of GaAs, which starts and ends with As atoms, and a block of Fe$_3$Si, which consists of 7 layers with alternate arrangement of mixed Fe-Si and pure Fe sheets.
Due to periodic boundary conditions, our model has to include two interfaces in the cell and effectively we obtain a multilayer structure. In order to have the same terminations on both sides of the blocks, we introduce odd numbers of layers in the GaAs and Fe$_3$Si blocks.
Other methods of calculations for polar GaAs surfaces involve constructing supercells with double slabs and additional techniques for reducing the charge transfer~\cite{Kaxiras1987,Ohno1990,Mankefors1999}.

Only As terminated GaAs layers were examined since this is the interface configuration of the samples used in experiments validating the calculations~\cite{Kalt2018}.
The Fe$_3$Si fragment can start either with a Fe-Si layer (variants A and C) or with a pure Fe layer (variants B and D). 
The variants A and C (as well as B and D) are distinguished by orientation of the Fe-Si layer in reference to the Ga and As layers (see Fig.~\ref{fig:structure}).
We adopt the notations of the variants introduced by Kaganer {\it et al.}~\cite{Kaganer2008}. In the present models, we do not take into account possible lattice deformations and crystal defects, which may exist in real samples.
Modeling of such effects requires larger supercells, which practically precludes phonon calculations. 

All studied Fe$_3$Si/GaAs heterostructures have tetragonal symmetry (space group $P\bar{4}2m$). 
The electronic structure was calculated using the primitive unit cells with a square base $a$ and height $c$ depicted in Fig.~\ref{fig:structure}. 
Phonon calculations were carried out in the supercell expanded to 2$\times$2$\times$1. 
The crystallographic lattice parameters are related to the lattice vectors: $\textbf{a}_\mathrm{cr} = \textbf{a}+\textbf{b}$, $\textbf{b}_\mathrm{cr} = \textbf{a}-\textbf{b}$,  $\textbf{c}_\mathrm{cr} = \textbf{c}$, where \textbf{a}, \textbf{b}, and \textbf{c} are lattice vectors of the primitive unit cell and $|\textbf{a}|=|\textbf{b}|$= $a$. 
In calculations periodic boundary conditions were imposed along all axes. 

\subsection{Computational procedure}
\label{sec:computational}

The spin-polarized DFT calculations were performed using the Vienna Ab initio Simulation Package (VASP)~\cite{VASP1,VASP2}. 
The exchange-correlation functional was calculated within the generalized gradient approximation (GGA) developed by Perdew, Burke, and Enzerhof (PBE)~\cite{PBE1,PBE2}.
The valence electrons (electron configurations: $d^7s^1$, $s^2p^2$, $s^2p^1$, and $s^2p^3$ for the Fe, Si, 
Ga and As atoms, respectively) were represented by plane-wave expansions with an energy cutoff of 400~eV. 
The wave functions in the core region were evaluated using the full potential projector augmented-wave method~\cite{PAW1,PAW2}.
Optimizations of the structural parameters and the electronic structure of the bulk Fe$_3$Si and GaAs were performed for primitive cells using a 24$\times$24$\times$24 Monkhorst-Pack grid of {\bf k}-points whereas a 16$\times$16$\times$4 grid was used in the case of the multilayer primitive cells.
The structures were optimized using the conjugate gradient technique with the energy convergence criteria
set at $10^{-7}$ and $10^{-5}$~eV for the electronic and ionic iterations, respectively.

Magnetic moments of the optimized structures were evaluated by summation of the spin densities within spheres containing particular atoms. 
The results were compared to the values calculated with the Bader method \cite{bader}.

For each relaxed structure, the phonon dispersion relations as well as the total and partial, element-projected PDOS were calculated using the direct method~\cite{PHONON1} implemented in the Phonon software~\cite{PHONON2}.
This method utilizes the Hellmann-Feynman forces acting on all atoms in a given supercell caused by single-atom displacement from its equilibrium position. 
In phonon calculations the {\bf k}-mesh was reduced to 4$\times$4$\times$1.

\subsection{Experimental details} 
\label{sec:nis}

The Fe$_3$Si(001)/GaAs(001) heterostructure with the Fe$_3$Si layer thickness of 3 monolayers (ML) was grown via molecular beam epitaxy~\cite{Herfort2008,Jenichen2014} using iron enriched to 96\% in the M\"{o}ssbauer active isotope $^{57}$Fe. 
A monolayer defined as 1\,ML = $0.5$ $a_{cr}$ consists of two atomic layers, one of them with iron atoms only and the other containing both Fe and Si atoms.
In Fe$_3$Si(001), 1\,ML amounts to a thickness of 0.28 nm, as deduced from the XRD study \cite{Herfort2004}.
By electron and x-ray diffraction and transmission electron microscopy the growth of epitaxial, strain-free two-dimesional Fe$_3$Si structures was confirmed. 
The sample was covered with 4 nm of amorphous Ge to prevent oxidation and to eliminate surface vibrational modes. 
The details of sample preparation and characterization were presented in the Supplemental Material of Ref.~\cite{Kalt2018}. 

The vibrational properties were studied via the Fe-partial PDOS~\cite{Kohn2000} 
obtained by nuclear inelastic scattering~\cite{Seto1995,Sturhahn1995} performed at 
the Dynamics Beamline P01~\cite{Wille2010} of PETRA III with an energy resolution 
of 0.9 meV in grazing-incidence geometry.
The wavevector of the x-ray beam was oriented along the [100] crystal direction of the Fe$_3$Si(001) film;
therefore, only phonon modes with nonzero components of the polarization vectors along this direction could be detected~\cite{Chumakov1999}. Consequently, one measures the Fe-partial PDOS, which is projected 
along the [100] direction (hereinafter referred to as "Fe$_3$Si[100]-projected, Fe-partial PDOS").

\section{Theoretical results}

\subsection{Structural properties}
\label{sec:theo_structural}

We started the calculations with optimization of lattice parameters of bulk Fe$_3$Si and GaAs. 
Obtained cubic lattice constants $a_{cr}$ are equal to $5.599\,$\AA~and $5.764\,$\AA~for the Fe$_3$Si and GaAs crystals, respectively.
These results are in good agreement with the experimental values of 5.665~\AA~\cite{exp_Fe3Si} and 5.653~\AA~\cite{exp_GaAs} deduced from the XRD studies.
The deviation of 1--2\% from the experimental lattice constants is typical for DFT calculations. 

Next, both the lattice parameters and atomic positions of the considered Fe$_3$Si/GaAs(001) multilayers  were calculated using the data obtained for bulk crystals as starting values. 
The distances between neighboring layers are initially set to the bulk values in the Fe$_3$Si and GaAs blocks (1.40~\AA~and 1.44~\AA, respectively) and to the average value of 1.42~\AA~at the interface.  
The structural data of the optimized heterostructures are presented in Table~\ref{tab_structure}. 
We observe a shortening of the in-plane lattice parameters $a$ and an elongation of interlayer distances.

\begin{table}[t] 
\caption{Structural parameters 
and formation energies ($E_{\mathrm{form}}$) calculated for the four variants of the Fe$_3$Si/GaAs heterostructure.} 
\begin{ruledtabular}
\begin{tabular}{lcccc}
                 & variant A   & variant B & variant C  &  variant D \\
\hline
\multicolumn{5}{c} {lattice parameters (\AA)} \\
$a$           &  3.883     &  3.938   &  3.837    &  3.904 \\
$a_{\mathrm{cr}}$    &  5.491     &  5.569   &  5.427    &  5.522  \\
$c $          & 18.646     & 17.538   & 19.042    & 18.066  \\
\multicolumn{5}{c} {thicknesses of blocks (\AA)} \\
Fe$_3$Si         &  8.795     &  8.509   &  9.178    &  8.749  \\
GaAs             &  6.299     &  6.271   &  6.450    &  6.257  \\
\multicolumn{5}{c} {distance between blocks (\AA)} \\
Fe$_3$Si$-$GaAs   &  1.776     &  1.379   &  1.705    &  1.530     \\
\multicolumn{5}{c} {distances between layers in the Fe$_3$Si block (\AA)} \\          
1st$-$2nd layers   &  1.434     &  1.425   &  1.480    &  1.522  \\
2nd$-$3rd layers   &  1.485     &  1.405   &  1.466    &  1.399  \\
3rd$-$4th layers   &  1.439     &  1.416   &  1.468     & 1.430  \\
\multicolumn{5}{c} {distances between layers in the GaAs block (\AA)} \\          
1st$-$2nd layers   &  1.640     &  1.649   &  1.698    &  1.630  \\
2nd$-$3rd layers   &  1.510     &  1.487   &  1.527    &  1.499  \\
\multicolumn{5}{c} {rumpling parameter in the Fe$_3$Si block (\AA)} \\
1st layer        &  0.080     &  0.017   &  0.288    & 0.048  \\
2nd layer        &  0.103     &  0.019   &  0.074    & 0.117  \\
3rd layer        &  0.004     &  0.008   &  0.003    & 0.000  \\
4th layer        &  0.000     &  0.000   &  0.000    & 0.000  \\
\hline
$E_{\mathrm{form}}$ (eV/atom)& -0.127   & -0.148   & -0.144    &-0.118 \\

\end{tabular}
\end{ruledtabular}
\label{tab_structure}
\end{table}

The thicknesses of Fe$_3$Si and GaAs blocks depend significantly on the structural variant, but they are always larger than the starting values of 8.40~\AA\ and 5.65~\AA, respectively. 
In turn, the distances between the blocks are close to starting values in variants B and D, while they are clearly larger in variants A and C.
This may be evidence of the repulsive interaction between neighboring Fe-Si and As layers. 
Further analysis demonstrates that the Ga-Si distance (3.69~\AA) in variant C is longer than the respective Ga-Fe distance (3.42~\AA) in variant A.
Similarly, the As-Si distance (3.13~\AA) in configuration D is longer than the As-Fe distance (2.82~\AA) in variant B. Thus the interaction between next nearest neighbors Ga-Si and As-Si is weaker than those between corresponding Ga-Fe or As-Fe atoms. 

The interactomic forces acting in the specific geometry of Fe$_3$/GaAs heterostructures generate only small structural changes perpendicular to the layers (the so-called  intralayer rumpling).
In Table~\ref{tab_structure}, the intralayer rumpling parameter, defined as the maximal difference between the $z$~coordinates of atoms belonging to one layer, is presented for each layer in the Fe$_{3}$Si block. 
The rumpling parameters in the first and second layers are much larger than in the third layer. In the fourth layer, the rumpling is zero due to the imposed symmetry. 
The largest displacement of atoms is found in the first layer of variant C, and this is the only case where rumpling is larger than in the second layer. 
Comparing the results for the first layers, larger rumpling parameters are found in variants A and C.
It can be explained by the presence of two types of atoms, Fe and Si,
which show significantly different shifts perpendicular to the interface due to interactions with the substrate. 
The distances between successive Fe$_3$Si layers and the rumpling in the first and second layers suggest that the relaxed configuration of variant B is the closest to the initial structure, while in the remaining variants the modification of the interface is much larger.

To compare the energetic stability of different interface configurations, the formation energies of the multilayers are calculated and presented in Table~\ref{tab_structure}. 
The formation energy $E_{\mathrm{form}}$ per atom of the Fe$_3$Si/GaAs supercell can be described as the difference between the total energy $E_{\mathrm{tot}}$ and chemical potentials of the constituents according to the formula
\begin{equation}
 E_\mathrm{form}= \frac{ E_\mathrm{tot} - nE_\mathrm{Si} - mE_\mathrm{Fe} - kE_\mathrm{Ga} - lE_\mathrm{As} }{ n+m+k+l } ,
\end{equation}
where $n$, $m$, $k$, and $l$ are the numbers of Si, Fe, Ga, and As atoms and $E_\mathrm{Si}$, $E_\mathrm{Fe}$, $E_\mathrm{Ga}$, and  $E_\mathrm{As}$ are the corresponding chemical potentials. 
The chemical potentials of the constituents were determined from calculations for each element in its native crystal form: 
iron in bcc ($Im\bar{3}m$), silicon in diamond ($Fd\bar{3}m$), arsenic in rhombohedral ($R\bar{3}m$), and gallium in centered orthorhombic ($Cmca$) structures. We obtained $E_\mathrm{Si}=-5.425$ eV/atom, $E_\mathrm{Fe}=-8.298$~eV/atom, $E_\mathrm{Ga}=-2.911$~eV/atom and $E_\mathrm{As}=-4.703$~eV/atom. 
The calculated binding energies are negative, indicating that the considered structures are energetically stable, in contrast to the artificial structures such as Fe/Au multilayers that require additional energy inserted to the system to be formed~\cite{Sternik2006}.
Although configurations B and C have lower formation energies than variants A and D, small differences between them suggest that all variants can be created. 

\subsection{Magnetic moments}
\label{sec:theo_magnetic}

\begin{figure}[t]
\centering
\includegraphics[width=\linewidth]{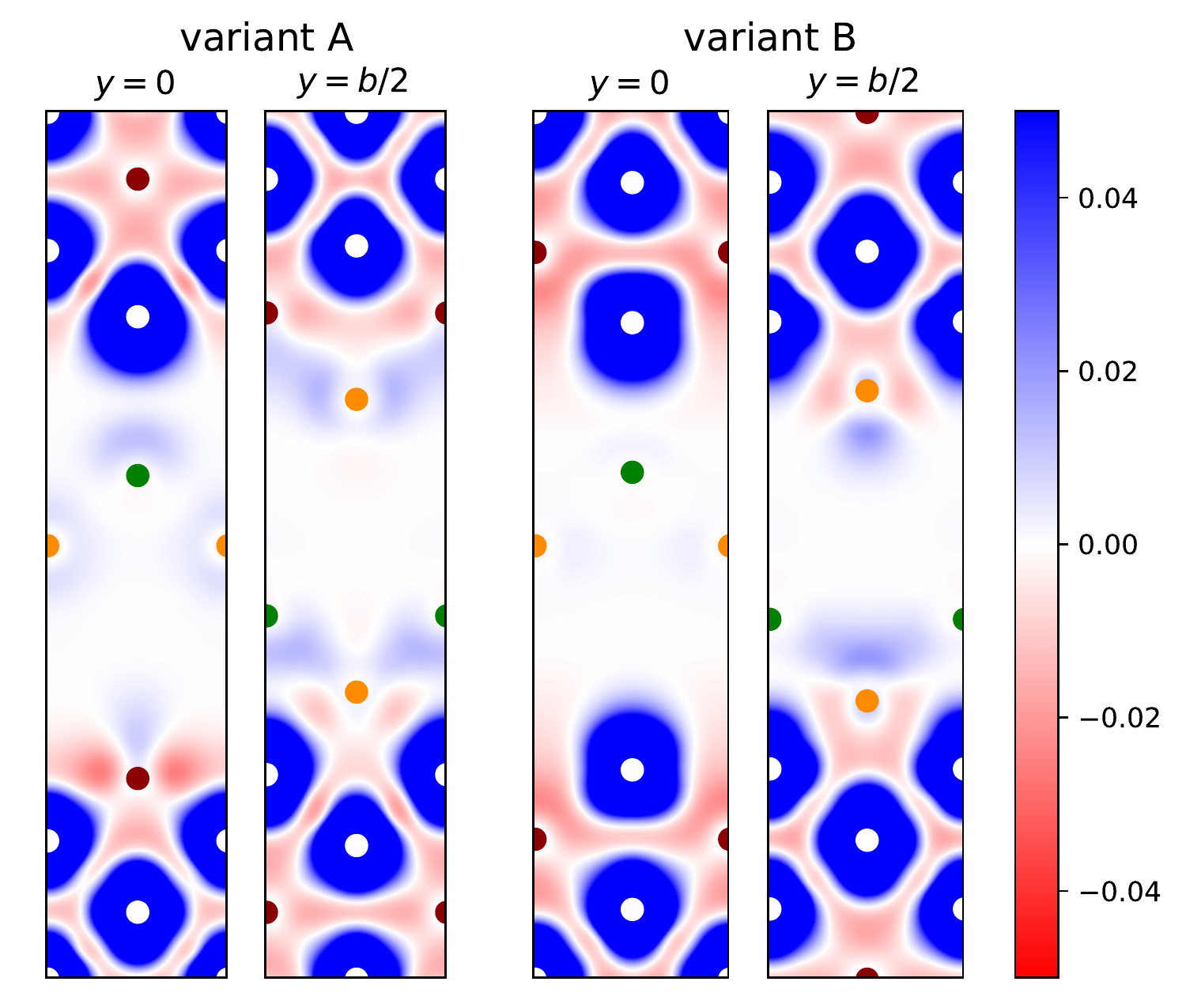}
\caption{\label{fig:chgdiff_AB} Maps of the magnetization density ($\mu_B$/\AA$^3$) for variants A and B. The slices are perpendicular to the \textbf{b} vector (y axis) with either $y=0$ or $y=b/2$. The positions of Fe, Si, As, and Ga atoms are marked by white, red, orange, and green dots, respectively.}
\end{figure}

The spin-polarized calculations are performed for the ferromagnetically ordered iron atoms. In Fig.~\ref{fig:chgdiff_AB}, we show the magnetization density obtained as a difference of the spin-up and spin-down electron density for variants A and B.  
Since we discuss mainly the differences between the mixed Fe-Si and pure Fe interface layer, we choose only variants A and B (this applies also to charge densities discussed in Sec.~\ref{sec:theo_charge}).
For each variant, two cuts are taken using slices perpendicular to the {\bf b} vector with either $y=0$ or $y=b/2$. 
These cuts ensure that the close vicinity of each atom is represented in the maps. Most of the magnetization density is found on the Fe atoms, therefore in order to present relatively small contributions from other atoms, we show a small range of densities around zero. 
To quantitatively analyze the impact of the interface on magnetic moments, we use two different methods of calculating their values. 

In the first approach, we integrate the difference between spin up and spin down density within spheres surrounding the Fe ions as a function of their radius.
The spin density around Fe ions is spherical to a good approximation (the blue regions in Fig.~\ref{fig:chgdiff_AB} show only the peripheral areas of the distribution)  and between the ions there are regions with opposite moment density. We therefore choose, for each nonequivalent Fe atom, the maximal value of the integral as an estimation of the magnetic moment (see Fig.~\ref{fig:cells}). 

We compared these results with the Bader analysis (originally invented to calculate atomic charges) and got a very good agreement: the magnetic moments on Fe atoms are lower only by 2--4\% with respect to the first method. Although the integration regions in the Bader analysis are not spherical, the differences occur in regions of low densities of magnetic moment and their contribution to the integral is small.
We conclude that the results in Fig.~\ref{fig:cells} reliably reflect the changes in the dominating magnetic ions in the vicinity of the interface.

The calculated magnetic moments of Fe ions in bulk Fe$_3$Si  are $2.58\,\mu_B$ in mixed Fe-Si and $1.34\,\mu_B$ in the pure Fe layer (all magnetic moments in the bulk are obtained using the integration within a sphere method, and their values are close to previously published results~\cite{Dennler2006}).
Comparison between the  magnetic moments on Fe atoms in heterostructures and the bulk Fe$_3$Si reveals the largest changes in pure Fe layers at the interfaces in variants B and D. In variant D, the magnetic moments increase by 66\% and 72\%, and in variant B by
16\% and 40\% in the two nonequivalent positions. In mixed Fe-Si interface layers, the Fe magnetic moments are enhanced only by 7\% (variant A) and 2\% (variant C).
The enhancement of magnetic moments is induced by the shift of the spin-down states to higher energies (discussed in Sec.~\ref{sec:theo_electron}). Therefore, their occupancy decreases and makes the total magnetic moment larger. Since almost all spin-up Fe states are occupied in the bulk and at the interface, their contributions to the magnetic moments are very similar.
A small enhancement of magnetic moments is found also in the second (pure Fe) layers of variants A and C.
In the third layer, magnetic moments are close to the bulk values, except for a small reduction found in variant D (10\%), where deviations from the bulk in the EDOS are more pronounced than in variant B. 

Magnetic moments on Si atoms (Fig.~\ref{fig:cells}) were obtained using an atomic radius corresponding to the maximal absolute value of the magnetization integrated over a sphere.
As can be seen from the spin density maps, areas surrounding Si atoms in Fe$_3$Si fragments include spin density of the opposite sign to the dominating Fe moments and the largest values can be found between ions. In the mixed Fe-Si layer adjacent to the interface (variant A), the spin density of Si atoms is strongly modified and includes regions with both positive and negative signs.  The integrated moment is decreased with respect to the bulk value -0.11~$\mu_B$ in variants A and C,  and slightly increased in variants B and D where Si atoms are in the second layer. In the third and fourth layers, the results are very similar to the bulk values.
The Bader method gives smaller magnetic moments ($-0.02$ or $-0.03$~$\mu_B$) for Si ions. This discrepancy can be expected since now the differences in the integration volumes contribute significantly to the result. As for the substrate layer with Ga and As ions, a small magnetic moment density is generated, but it is difficult to determine spheres around ions and the Bader analysis gives results below 0.1~$\mu_B$.

\begin{table}[t] 
\caption{Magnetic anisotropy energy (MAE), in meV per primitive unit cell, calculated for different Fe$_3$Si/GaAs interface variants and compared with the values obtained for cubic and tetragonally deformed Fe$_3$Si crystal. The total energies are calculated for spins aligned along various directions in the crystallographic cell of Fe$_3$Si.}
\begin{ruledtabular}
\begin{tabular}{ccccccc}
      MAE                 & \multicolumn{4}{c}{interface variants} & \multicolumn{2}{c}{bulk Fe$_3$Si} \\
   (meV/p.u.c)            & A       & B        & C       &  D     &cubic & tetragonal\\
\hline
$E_{[100]}- E_{[001]}$    &  0.10   & 0.17     & 0.10    & -0.57   &  0.00 & -0.26 \\
$E_{[110]} - E_{[001]}$   &  -9.28  &  -7.38   & -8.03   & -1.25   &  0.01 & -0.27 \\
$E_{[110]} - E_{[100]}$   &  -9.38  &  -7.55   & -8.13   & -0.68   &  0.01 & -0.01 \\
\end{tabular}
\end{ruledtabular}
\label{tab_mae}
\end{table}

Finally, we discuss the magnetic anisotropy in the studied heterostructures. 
The magnetic anisotropy energy (MAE) can be obtained as a difference in the total energy between systems with magnetic moments ordered along selected different directions. 
In a perfect cubic Fe$_3$Si crystal, the preferred magnetization direction (easy axis) is one of the main crystallographic axes~\cite{Lenz2005}.
Our calculations confirm that magnetization along the [100] or [001] direction of the cubic crystal gives lower energy than magnetization along the diagonal [110] direction (see Table~\ref{tab_mae}).
In a thin Fe$_3$Si film placed on a substrate, magnetic anisotropy can arise due to symmetry breaking at the interface and/or the interatomic coupling between the Fe$_3$Si layers and the substrate.
The calculations carried out for the Fe$_3$Si/GaAs multilayers unveil that the [110] direction of the crystallographic unit cell is the preferred magnetization direction in all variants, although in variant D the energy differences strongly deviate from values obtained for the other heterostructures (Table~\ref{tab_mae}).  
This result shows a very good agreement with the measurements for ultrathin epitaxial Fe$_3$Si films on GaAs(001), which revealed the easy axis pointing along the [110] direction~\cite{Herfort2008}. 
To verify if such spin alignment can be caused by a tetragonal  distortion present in the Fe$_3$Si/GaAs heterostructures, we performed additional calculations for the Fe$_3$Si bulk crystal with $c/a=1.1$.
We found that the tetragonal  deformation results in the easy axis orientation along either the [110] or [100] directions as the MAE is almost the same in both cases (Table~\ref{tab_mae}).
This implies that the geometric distortion on its own is sufficient to align the spins in the $xy$ plane, but it does not explain the differences between $E_{[110]}$ and $E_{[100]}$  obtained for the Fe$_3$Si/GaAs heterostructures (the last line of Table~\ref{tab_mae}). 
This result suggests that the vicinity of the GaAs substrate has a crucial impact on the spin alignment in the Fe$_{3}$Si thin film.

\subsection{Charge density distribution}
\label{sec:theo_charge}

\begin{figure}[t]
\centering
\includegraphics[width=\linewidth]{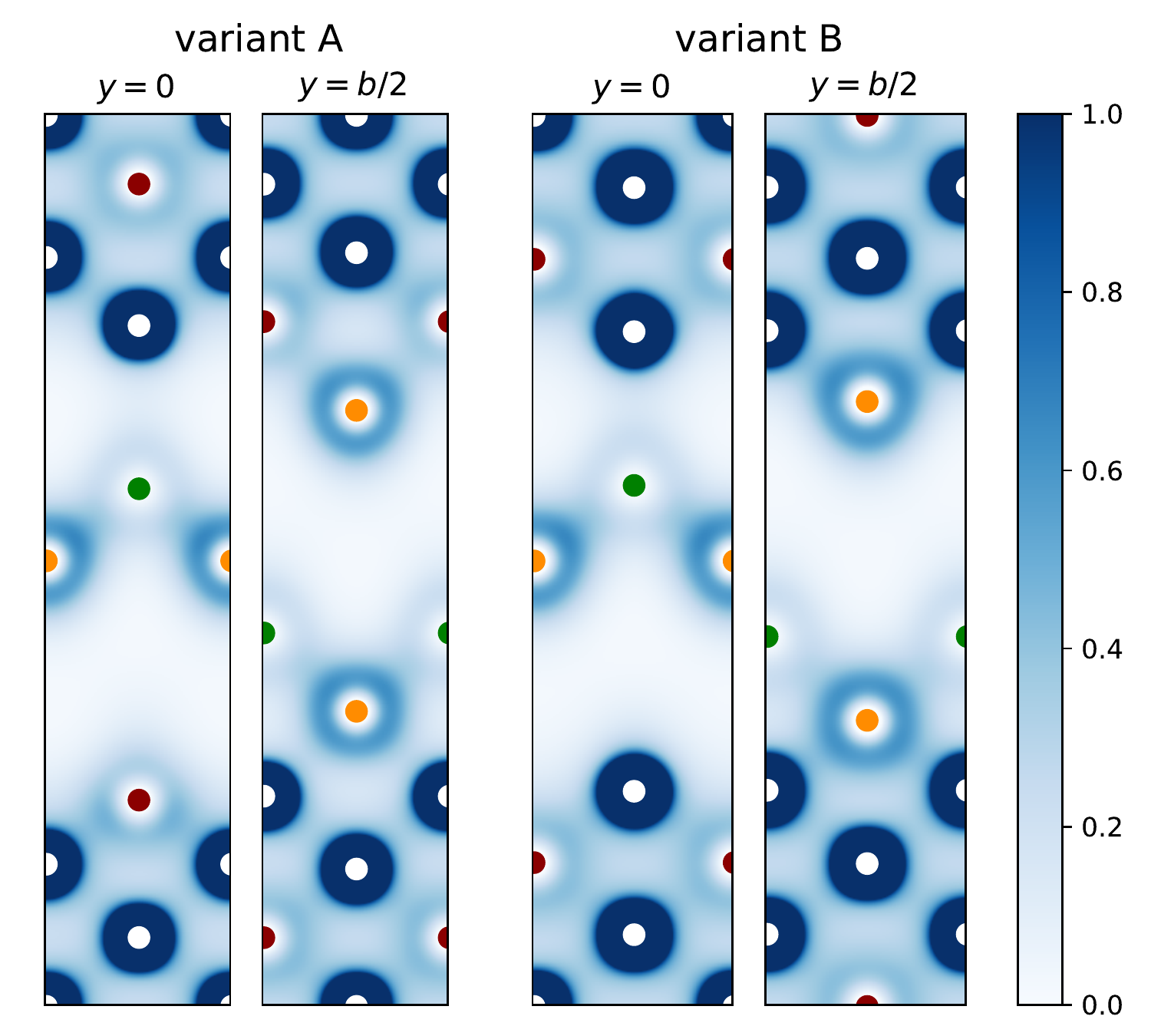}
\caption{
\label{fig:charge_AB}
Maps of the charge density (e/\AA$^3$) for variants A and B. The slice orientations and colors marking atomic positions are the same as in Fig.~\ref{fig:chgdiff_AB}.} 
\end{figure}

\begin{figure}[t]
\centering
\includegraphics[width=0.8\linewidth]{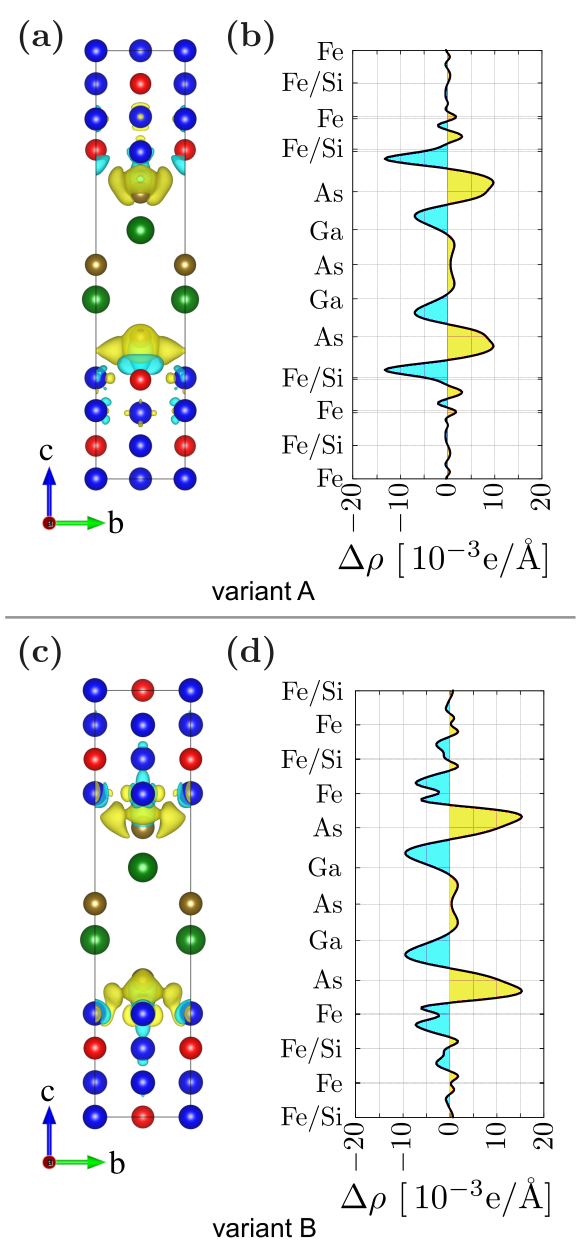}
\caption{
\label{fig:diff_chg_AB}
Charge density differences $\Delta \rho$ (left panels) and corresponding plane-averaged charge density difference $\Delta \rho (z)$ in the $z$-direction (right panels) for variants A and B (top and bottom panels, respectively).
Yellow and cyan regions indicate the electron accumulation and depletion, respectively.
}
\end{figure}

The charge density distribution is presented in the form of two-dimensional maps in Fig.~\ref{fig:charge_AB}. 
To make small charge densities in the interatomic space visible the upper limit of the color map is fixed to a low value.   
Two regions with charge density distributed in apparently different ways can be distinguished in Fig.~\ref{fig:charge_AB}. 
In the GaAs part, electrons congregate mainly around the atoms and in Ga-As bonds while in the Fe$_3$Si layer electrons fill also the interatomic space indicating a metallic state.
The higher electron density observed between interface As atom and neighboring Fe or Si atoms suggests that the As-Si and As-Fe bonds are formed through the sharing of valence electrons.

\begin{figure}[t]
\centering
\includegraphics[width=\linewidth]{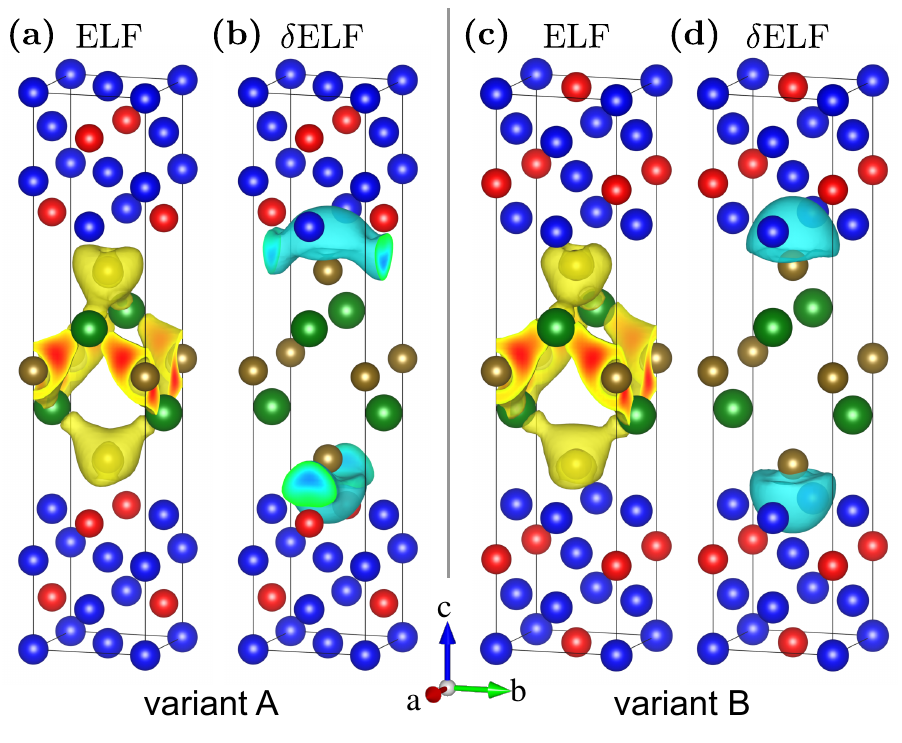}
\caption{
\label{fig:elf_AB}
(a),(c) Isosurface of electron localization functions ELF~=~0.55. (b),(d) Differences ($\delta$ELF) between ELFs of the interface and the component slabs. 
Blue surfaces of $\delta$ELF indicate the reduction of ELF values.
Atoms colors and cells orientation are the same as in Fig.~\ref{fig:cells}.
}
\end{figure}

The changes in charge distribution induced by the Fe$_3$Si/GaAs interface can be specified more precisely using three-dimensional plots of the difference between the total charge density of the interface ($\rho_{\mathrm{FeSi/GaAs}}$) and charge densities of systems with vacuum substituting for GaAs ($\rho_{\mathrm{FeSi/vac}}$) or Fe$_3$Si ($\rho_{\mathrm{vac/GaAs}}$) in accordance with the formula:
\begin{eqnarray}
\Delta \rho = \rho_{\mathrm{FeSi/GaAs}} - \rho_{\mathrm{FeSi/vac}} - \rho_{\text{vac/GaAs}}.
\label{eq:rho}
\end{eqnarray}
The charge rearrangement for variants A and B is illustrated in Figs.~\ref{fig:diff_chg_AB}(a) and \ref{fig:diff_chg_AB}(c) with the yellow regions representing charge accumulation and the light blue regions indicating charge depletion.
It is clearly visible that most significant charge redistribution takes place in the close vicinity of the metal-semiconductor contact where electron transfer from metallic Fe$_3$Si to semiconducting GaAs is observed. 
In other Fe$_3$Si layers, the charge redistribution is very small. 

\begin{figure*}[!t]
\centering
\includegraphics[width=\linewidth]{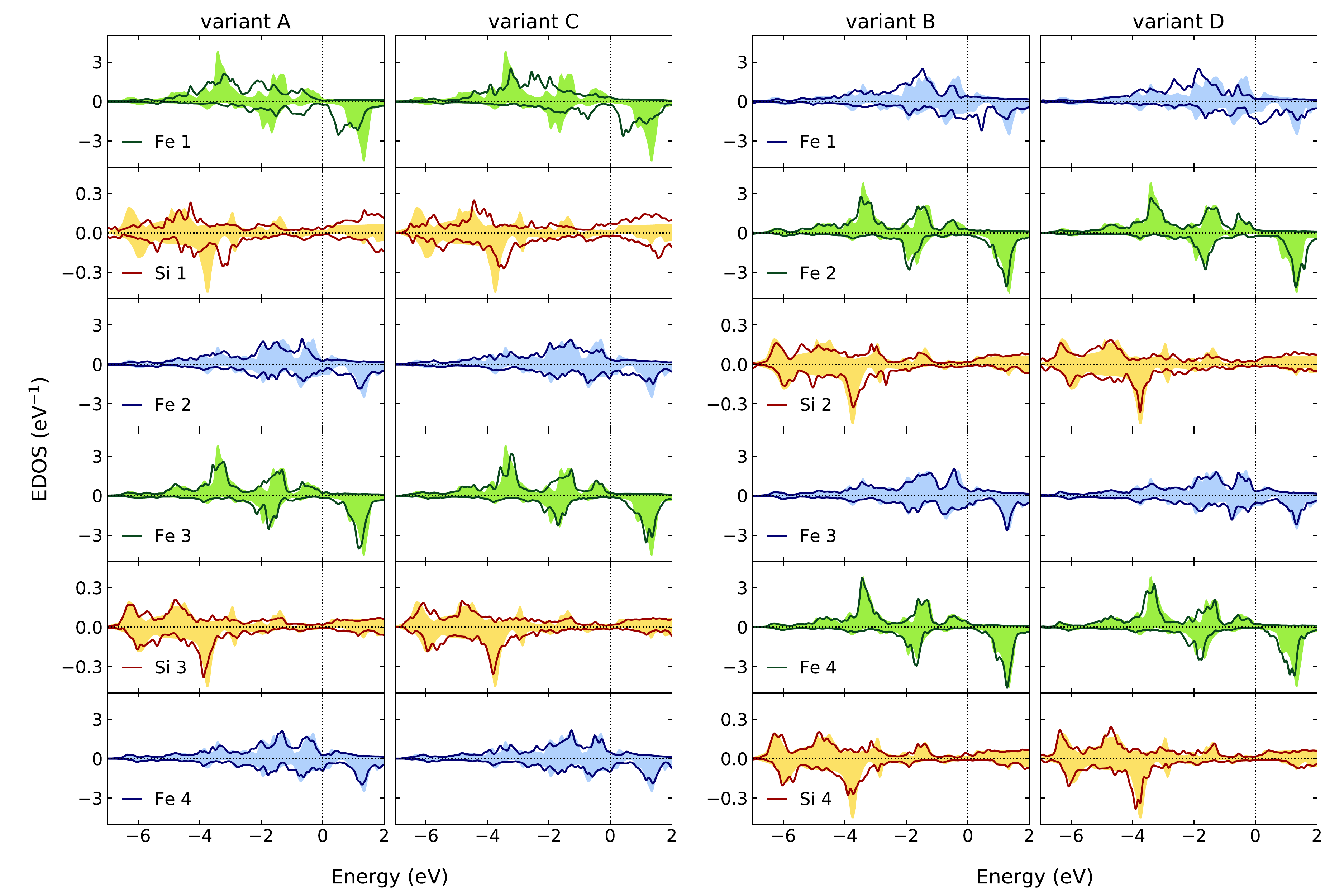}
\caption{\label{fig:eldos_AC} Electron density of states calculated for each layer in variants A, C (left panels) and B, D (right panels) compared with the corresponding results for the bulk Fe$_3$Si crystal (filled area). For mixed Fe/Si layers, the densities for Fe and Si atom are plotted separately (green and red lines, respectively), while for pure Fe layers the averaged electron density is shown (blue lines).   
}
\end{figure*}

The charge transfer across the interface can be evaluated quantitatively by the plane-averaged charge density difference
$\Delta\rho$ obtained for each $z$~\cite{min.park.17}.
The calculated $\Delta\rho (z)$ are plotted for variants A and B in Fig.~\ref{fig:diff_chg_AB}(b)  and (d), respectively.
The positive values of $\Delta\rho (z)$ represent the electron accumulation while the negative ones -- depletion.
We can see that electron density which is moved towards the As atom originate either from the first Fe$_3$Si layer or from neighboring Ga atoms. In variant A, electrons are mainly transferred from the side of the Fe-Si layer which is in direct contact with GaAs. 
In contrast, in variant B the relocation of electrons from both sides of the Fe layers is observed. 
By integration of the charge differences presented in Figs.~\ref{fig:diff_chg_AB}(b) and \ref{fig:diff_chg_AB}(d), we found that in both variants a small fraction of electron density is transferred from Fe$_3$Si to GaAs.
The charge transfer is a typical phenomenon for metal/semiconductor interfaces and it results in the interface dipole formation~\cite{tung.01,min.park.17}.

Apart from the charge redistribution, also the character of electronic states in the contact area may be modified. It can be studied by calculating the electron localization function (ELF)~\cite{becke.edgecombe.90,savin.jepsen.92}, which takes values from 0 to 1.
Large values of ELF (close to 1) indicate a high probability of finding an electron at a given point and correspond to highly localized states, while smaller values are related to more itinerant/delocalized electrons. 
Figures~\ref{fig:elf_AB}(a) and (c) show isosurfaces for ELF equal to 0.55 calculated for variants A and B, respectively.
In both cases, the largest values of ELF (from 0.55 to 0.77) are found on As-Ga bonds in the middle of the GaAs block.
In the metallic Fe$_3$Si region, we find itinerant electrons with ELF below 0.4.

Analysis of the difference in ELF ($\delta$ELF) between the interface and the 
component slabs [defined analogously to $\Delta \rho$ in Eq. (\ref{eq:rho})] 
reveals a reduction of electron localization around the interface [Fig.~\ref{fig:elf_AB}(b) and (d)].
Different shapes of $\delta$ELF result from the shapes of ELF at the surface of the Fe$_3$Si slabs in variants A and B. 
At the GaAs surface, we observe an enhancement of electron localization close to the As atoms on the side of vacuum  (see Supplemental Material~\footnote{See Supplemental Material at [URL will be inserted by publisher] presenting the ELF of free-standing Fe$_3$Si and  GaAs slabs in variants A and B.}).
Similar enhancement is found in Fe$_3$Si only if the surface includes Si atoms (variant A).
When the Fe$_3$Si/GaAs interface is created, we observe a reduction of electron localization 
around As and Si.
In variant A, it is an elongated region placed close to As and Si [Fig.~\ref{fig:elf_AB}(b)], while in variant B, there is only spherical region around As [Fig.~\ref{fig:elf_AB}(d)].
This illustrates a change of the electronic states of the semiconducting elements from localized to more itinerant 
through formation of new bonds between slabs.

\subsection{Electron density of states}
\label{sec:theo_electron}

To investigate the influence of the GaAs substrate on the electronic states in the Fe$_3$Si block, we calculated the electron densities of states (EDOS) projected on Fe and Si atoms belonging to different layers. 
In Fig.~\ref{fig:eldos_AC}, the results obtained for all variants of the multilayers are compared with the EDOS of the Fe and Fe-Si layers in bulk Fe$_3$Si. 
In all variants, the largest differences are found in the first layer (Fe1, Si1) being in contact with GaAs, which modifies the bonding geometry and orbital hybridizations.
The changes occurring in the vicinity of the Fermi energy are mainly connected with the modification of the Fe spin-down states.
It results in a larger EDOS, compared with the bulk Fe$_3$Si, especially in the Fe1 layers in variants B and D.
The occupied spin-up states are more strongly modified in the Fe1-Si1 layer (A and C) than in the Fe1 layer (B and D). 
Comparing the electronic bands in the second layers, more pronounced changes are observed for the Fe2 layer in variants~A and~C.
Starting from the third layer, the differences between the bulk and heterostructures are very small. 

The stronger changes observed in variants A and C compared to variants B and D can be also connected with the modifications of the crystal structure in these configurations. As we discussed in Sec.~\ref{sec:theo_structural}, the presence of the mixed Fe1-Si1 layer generates larger deformations than in the monoatomic Fe1 layer. 
In variant B, we observe the weakest modification of the crystal structure, which shows the smallest rumpling parameters in the first and second layer. As a consequence, its electronic structure resembles the bulk EDOS even in the first Fe1 layer.

The observed changes in the electronic states at the interface influence the spin polarization.
We calculated the spin polarization at the Fermi energy for all variants and each layer of the Fe$_3$Si block using the formula
\begin{equation}
P=\frac{N_{\uparrow}(E_F)-N_{\downarrow}(E_F)}{N_{\uparrow}(E_F)+N_{\downarrow}(E_F)},
\end{equation}
where $N_{\uparrow}(E_F)$ and $N_{\downarrow}(E_F)$ are the spin-up and spin-down EDOS at the Fermi energy, respectively.
The results presented in Table~\ref{tab_pol} demonstrate large differences between spin polarization of different variants. 
As in the bulk crystal, the total polarization is negative in all variants. 
The largest values, even exceeding the spin polarization of the bulk crystal, are found in variants A and D.
In all layers, the spin polarization adopts the same sign as in the bulk, except for the first layer of variant A.

\begin{table}[b!]
\caption{Spin polarization at the Fermi energy for each layer and the total value at the interface. In bold we denote values for pure Fe layers (bulk value $-0.53$) and the remaining values are for the mixed
Fe-Si layers (bulk value $0.48$). The total polarization in the bulk Fe$_3$Si is equal to $-0.41$.}
\begin{ruledtabular}
\begin{tabular}{ c c c c c  }
layer  & variant A & variant B & variant C & variant D \\ [1.0ex]
\hline  
 1   &   -0.60     &       {\bf -0.44}        &       0.21         &      {\bf -0.76}        \\
 2   &    {\bf -0.51}    &        0.32        &      {\bf -0.38}          &          -0.02            \\
 3   &     0.14   &         {\bf -0.15}          &    0.26          &        {\bf -0.24 }            \\
 4   &     {\bf-0.69}    &         0.72          &     {\bf -0.39}         &        0.61             \\
\hline 
total Fe$_3$Si &     -0.52    &        -0.23         &     -0.26        &        -0.49           \\
\end{tabular}
\end{ruledtabular}
\label{tab_pol}
\end{table}

\begin{figure*}[t!]
\centering
\includegraphics[width=\hsize]{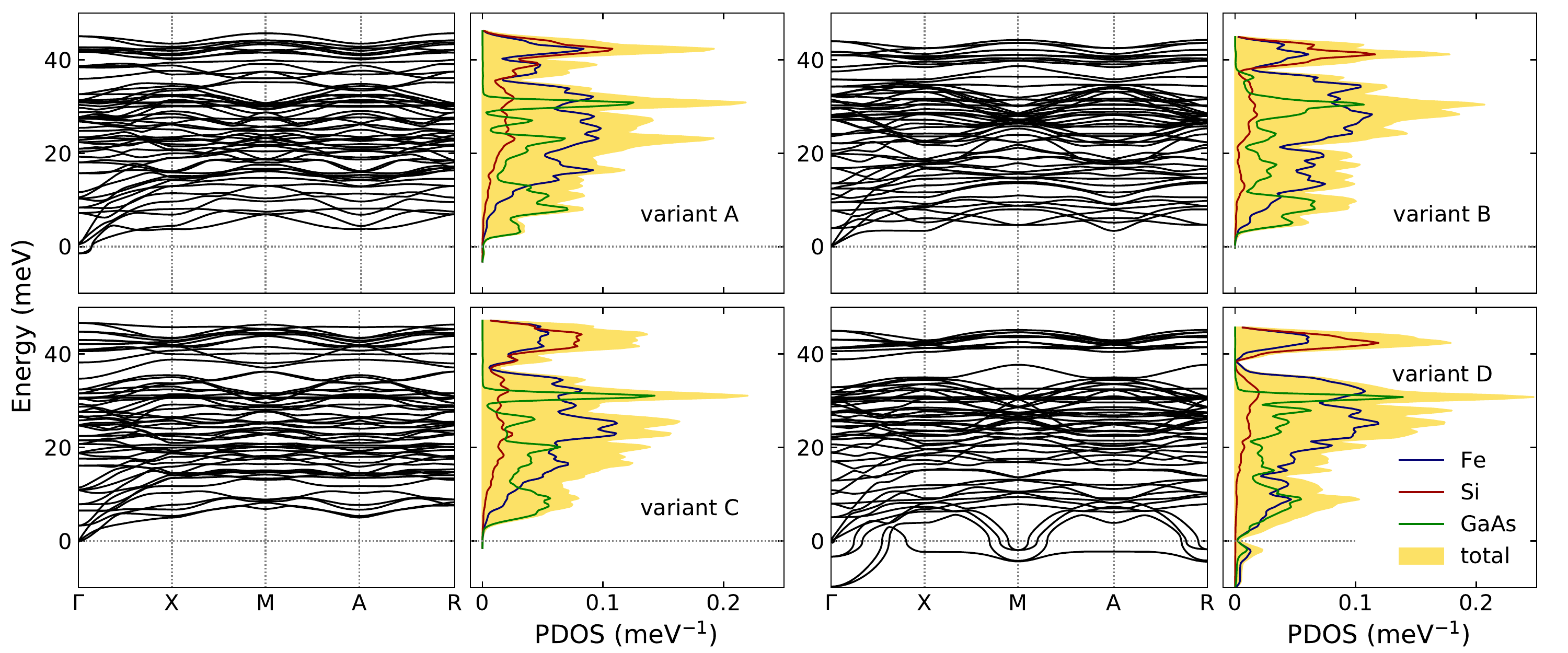}
\caption{\label{fig:fonony} Phonon dispersion relations and phonon density of states calculated for four Fe$_3$Si/GaAs heterostructures (variants A-D). The density of states is separated into contributions from Fe atoms (blue), Si atoms (red), and GaAs layers (green). The total phonon density of states is shown in yellow. }
\end{figure*}

The highest value of the spin polarization in the Fe1 layer of variant D is directly related to the largest spin-down EDOS at $E_F$. As we discuss in the next section, variant D is dynamically unstable, therefore the obtained EDOS and spin polarization
may be modified when the soft modes are stabilized in a real system. 

Highly polarized electron states in variant A, one of the configurations studied experimentally~\cite{Kaganer2008},
are very favorable for spintronic applications. The calculated spin-polarization 
is larger than the bulk value by 27\%. It is also larger than the previous theoretical results obtained for Fe$_3$Si/GaAs(110) multilayers~\cite{Grunebohm2010}. 
This spin-polarization enhancement agrees well with the measurements of spin-polarized currents in Fe$_3$Si/GaAs(001) thin films \cite{Ionescu2005}.

\subsection{Lattice dynamics}
\label{sec:theo_dynamic}

In this section, we study the lattice dynamics of the Fe$_3$Si/GaAs multilayers in the four atomic configurations shown in Fig.~\ref{fig:structure}. 
In our previous work \cite{Kalt2018}, we have presented solely the results obtained for variant C.
In Fig.~\ref{fig:fonony}, the calculated phonon dispersion relations and phonon density of states (PDOS) are presented. 
Dynamical stability (absence of imaginary modes) is only achieved in case of variant B with a pure Fe layer adjacent to the GaAs surface.
In variants A and C, a minor softening of the acoustic branches close to the $\Gamma$ point and along the $\Gamma$--Z direction (not shown) is found. 
However, these instabilities are very weak and their effect on the PDOS is negligible. They are observed mainly at wave vectors close to the $\Gamma$ point, which correspond to phonons with long wavelengths. Therefore, the origin of these soft modes may be connected with the limitation of the long-range interactions due to the supercell size.
Variant D shows the strongest instability with the imaginary modes existing in the whole Brillouin zone.
These unstable modes are localized mostly in the first Fe1 layer and the GaAs substrate.
We note a correlation between the dynamical stability and the formation energies discussed in Sec.~\ref{sec:structural}:
the largest and smallest binding energies
correspond to the most stable (B) and the most unstable (D) variants, respectively.

\begin{figure*}[t]
\centering
\includegraphics[width=\linewidth]{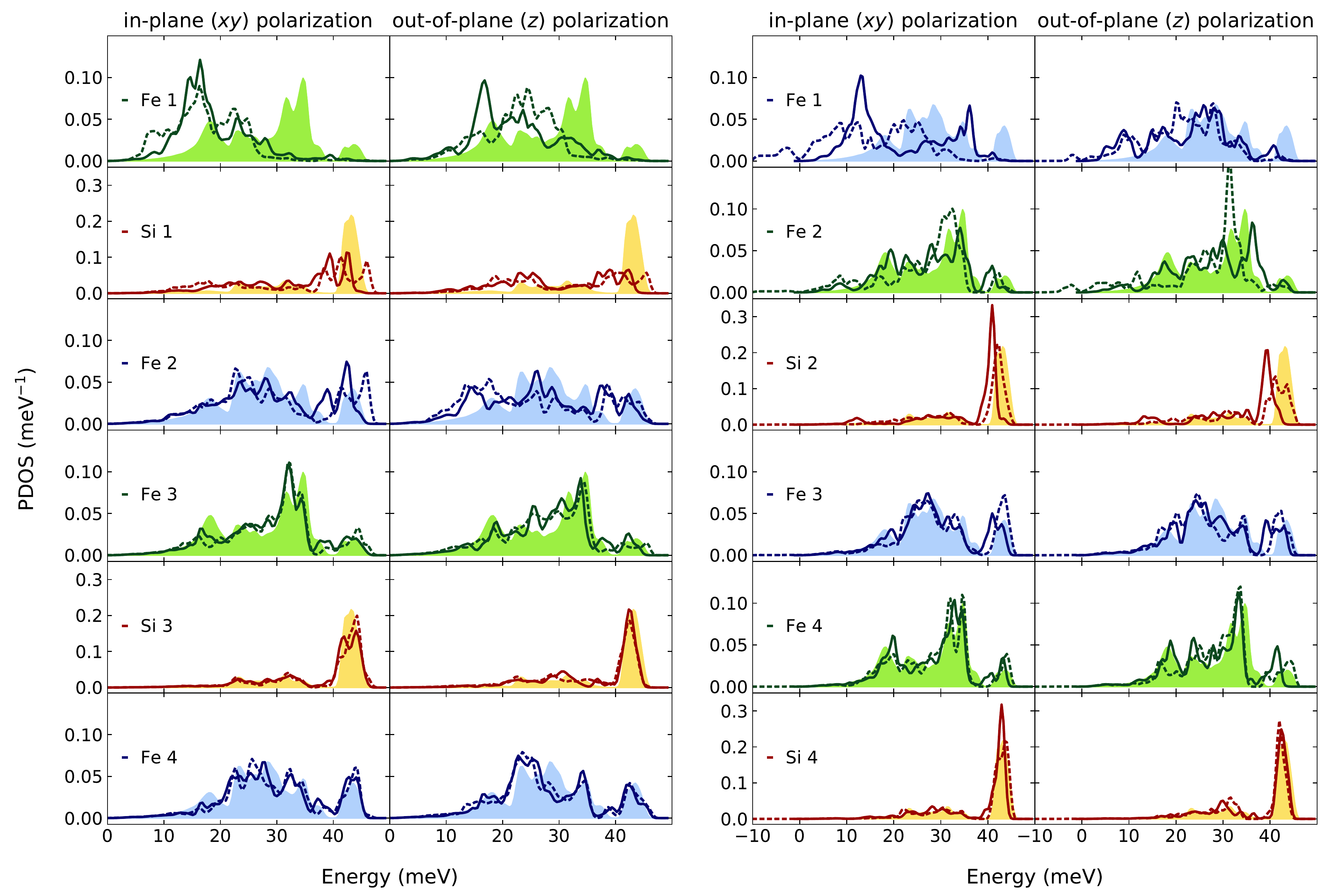}
\caption{\label{fig:pdos_AC} Phonon densities of states for variants A (solid line) and C (dashed line) presented in left panels and variants B (solid line) and D (dashed line) in right panels
projected onto layers and compared with the results for the bulk Fe$_3$Si crystal (shaded area). 
}
\end{figure*}

In the right panels of Fig.~\ref{fig:fonony}, the total PDOS together with the partial contributions of Fe, Si, and GaAs vibrations are shown. 
The characteristic features of the spectra obtained for different variants are very similar. 
The high energy limit of the spectra is about 45~meV, and the vibrations of Si and Fe atoms contribute to the high energy peaks around 40~meV, clearly separated from the remaining part of the spectra. 
In the medium part, the Fe vibrations dominate, although the highest intensity peak in this range is related to the GaAs vibrations.  
The low energy part, 0--12~meV, is dominated by vibrations of atoms in the GaAs layer and to a lesser extent by vibrations of the Fe atoms. 
In variant D, both Fe and GaAs vibrations soften reaching imaginary values. 

In order to elucidate contributions from atoms with different distances to the interface, the atomic and layer specific phonon spectra were also calculated.
The results are divided into two groups depending on the sequence of Fe and Fe-Si layers. 
In the left panels of Fig.~\ref{fig:pdos_AC}, the PDOS calculated separately for Fe1-Si1, Fe2, Fe3-Si3, and Fe4 layers of variants A (solid lines) and C (dashed line) are compared with spectra calculated for the Fe-Si and pure Fe sheets of the bulk Fe$_3$Si crystal (shaded areas). 
Additionally, two polarizations of vibrations are considered: in-plane $xy$ polarization and out-of-plane $z$ polarization are shown in the first and second columns, respectively.  

In general, the changes in the energy distributions in variant A are similar to those in variant C.  
The largest differences are observed in the Fe1-Si1 layer adjacent to the GaAs surface.
The Fe PDOS for both polarizations is notably shifted to lower energies.
For the in-plane vibrations, this effect is stronger in variant C, while the out-of-plane PDOS shows larger changes in variant A.
The most intense peaks of Si bulk spectrum, located around 42~meV, spread into a wider range of energies, but their shifts are not as large as in the Fe PDOS. 
Most likely this results from the stronger coupling between the Fe and GaAs modes at low energies.

In the second layer Fe2, the PDOS resembles that of the bulk crystal, however, clear differences are still observed. For example, the PDOS of the out-of-plane vibrations is largely enhanced below 20~meV.
The spectra of the layers further off the interface progressively approach the shape of the bulk PDOS.

In the same way, the PDOS for Fe1, Fe2-Si2, Fe3, and Fe4-Si4  layers of variants B (solid lines) and D (dashed lines) are compared in the right panels of Fig.~\ref{fig:pdos_AC}. The softening of the Fe1 vibrations in variant D leads to imaginary frequencies and therefore to dynamical instability of this heterostructure. This property is not transferred to the further layers in the case of in-plane vibrations, however, small contributions of imaginary frequencies are observed in Fe1 and Fe2 spectra of the out-of-plane vibrations.  The spectra of the third and fourth layers are  close to those found in the bulk crystal.

\section{Experimental results}
\label{sec:experimental}

\begin{figure}[!t]
\includegraphics[width=0.85\linewidth]{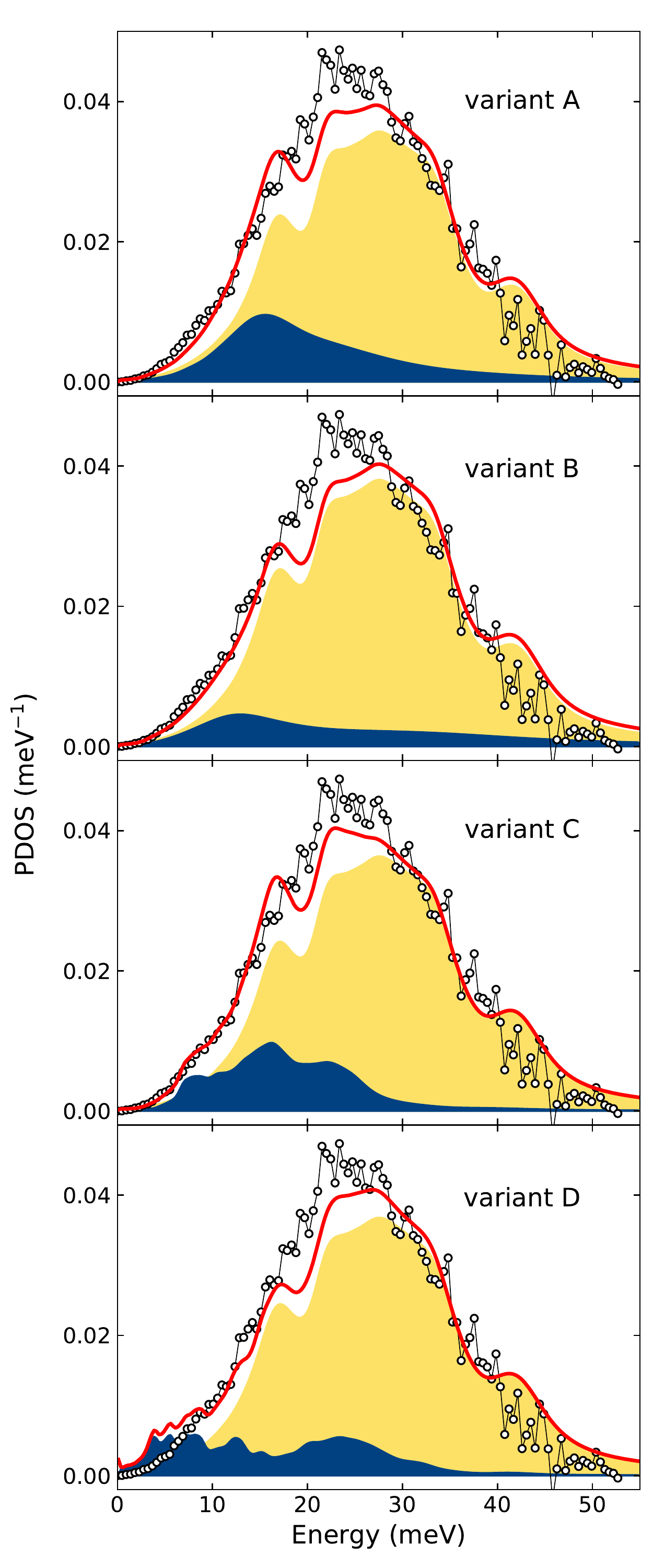}
\caption{Experimental Fe$_3$Si[100]-projected, Fe-partial PDOS of the 3\,ML sample (open circles) fitted with the model function g$_\mathrm{theor}(E)$ (red solid line) of the indicated interface variants. The yellow areas represent the bulk contribution $g_\mathrm{bulk}$ and the blue areas stand for the interface contribution $g_\mathrm{if}$ in the model function $g_\mathrm{theor}(E)$.
}
\label{fig:exp_pdos}
\end{figure}

The Fe$_3$Si/GaAs(001) heterostructure offers an 
opportunity to  compare the \textit{ab initio} calculated interface PDOS with experimental results. It can be grown almost strain-free with very high interface quality.
Furthermore, the outstanding sensitivity of the nuclear inelastic scattering enables the measurement of the PDOS in ultrathin films to experimentally confirm the effects described in Fig.~\ref{fig:pdos_AC}, which are strongly localized to almost a single atomic layer.
One problem that should be addressed is the c($4\times4$) reconstruction 
of the GaAs(001) surface, which is neglected in the theoretical model. However, there is no experimental evidence that the reconstruction is preserved when the GaAs(001) surface is covered by Fe$_3$Si.  
When an initially reconstructed GaAs(001) surface is overgrown with Fe$_3$Si, the epitaxial Fe$_3$Si/GaAs(001) interface 
is established, which enables the formation of crystalline Fe$_3$Si. 
Therefore, the dangling bonds on the GaAs(001) surface that induce the reconstruction are not present in the system anymore.
Additionally, the good agreement between the theoretical and experimental phonon spectra
as well as the thermoelastic properties obtained previously \cite{Kalt2018} justify the use of the unreconstructed Fe$_3$Si/GaAs($1\times1$)(001) interface in the theoretical model.

The Fe-partial interface specific \textit{ab initio} calculated PDOS for all four variants are compared with the experimental results for a 3\,ML Fe$_3$Si film epitaxially grown on the GaAs(001) substrate (Fig.~\ref{fig:exp_pdos}).
Despite the high sample quality, a broadening of phonon peaks caused by vibrations of atoms located at irregular sites has to be taken into account.
This is done by convolution of the theoretical PDOS with the damped harmonic oscillator (DHO) function characterized by a quality factor $Q$ \cite{Faak}. This introduces an energy-dependent broadening of the spectral features, with $Q$ being inversely proportional to the strength of the damping. To model the experimental data with the theoretical results we used the function $g_\mathrm{theor}(E)$ defined as follows:
\begin{equation}
g_\mathrm{theor}(E)=A\,g_\mathrm{if}(E,Q_\mathrm{if})+(1-A)g_\mathrm{bulk}(E,Q_\mathrm{bulk}),
\label{eq1}
\end{equation}
with $g_\mathrm{if}(E,Q_\mathrm{if})$ and $g_\mathrm{bulk}(E,Q_\mathrm{bulk})$ being the calculated $xy$-projected PDOS of the first interface layer (Fe1) of variants A-D and bulk Fe$_3$Si, respectively, convoluted with the DHO function with quality factors $Q_\mathrm{if}$ and $Q_\mathrm{bulk}$.
The parameter $A$ stands for the fraction of the interface Fe1 atoms in the sample. The value for $A$ estimated from the high resolution transmission electron microscopy study is 0.14 (for details see \cite{Kalt2018}). Furthermore, due to the grazing incidence geometry of the experiment, only the in-plane PDOS is measured, which allows for a comparison with the calculated $xy$-projected PDOS.
The experimental PDOS was compared to the function $g_\mathrm{theor}(E)$ with $Q_\mathrm{if}$, $Q_\mathrm{bulk}$, and \textit{A} being variable parameters. Analysis of the results using different numbers of atomic layers in $g_\mathrm{if}(E)$ (i.e., Fe$_1$, Fe$_1$\,+\,Fe$_2$, etc.) showed, that for all four variants the best results are obtained by taking only the first interface layer (Fe1) into account.

In Fig.~\ref{fig:exp_pdos}, a comparison of the obtained $g_\mathrm{theor}(E)$ for the different variants of interfaces with the experimental PDOS is displayed (red solid curve), while in Table~\ref{tab:parameters} the results of the least squares optimization  are given. $Q_\mathrm{bulk}$ is the same for all four types of interfaces and the relative bulk contribution to $g_\mathrm{theor}(E)$ varies between 83\% and 89\%. On the other hand the quality factors for the relative interface contribution differ significantly. For variants A and B a very strong damping with $Q_\mathrm{if}$ values of 2.5 and 1.5, respectively, is needed in order to obtain an optimum agreement with the experimental data. This results in an essentially  featureless interface PDOS (blue shaded area in Fig.~\ref{fig:exp_pdos}), which fails to describe the experimental low energy part between 5 and 10~meV. For variants C and D, $Q_\mathrm{if}$ is relatively high and the interface contribution exhibits pronounced features. In the case of variant D, the imaginary modes lead to  nonphysical phonon states at 0~meV. For variant C, the features entail a remarkable agreement between model and experiment. In general, variant C leads to the lowest residual sum squared (values in Table~\ref{tab:parameters} are normalized to variant C) and yields the best agreement with the expected \textit{A} value of 0.14~\cite{Kalt2018}. The deviations between experiment and theory observed for all variants in the range 17--27~meV may arise from the additional phonon modes induced by the Ge/Fe$_3$Si interface that is not accounted for by the model. 

\begin{table}[t]
 \caption{Modeling parameters obtained by fitting the experimental data with Eq. (\ref{eq1}), bulk quality factor Q$_\mathrm{bulk}$, interface quality factor Q$_\mathrm{if}$, relative interface contribution $A$, and the residual sum squared (rss) normalized to the value of variant C.
} 
\renewcommand{\arraystretch}{1.2}
\begin{ruledtabular}
 \begin{tabular}{c c c c c}
               & \hspace{1.5mm} Q$_\mathrm{bulk}$ \hspace{1.5mm} & \hspace{2mm}  Q$_\mathrm{if}$ \hspace{2mm}   &   \hspace{2.5mm}  $A$ \hspace{2.5mm}        &  \hspace{1mm}  norm. rss   \hspace{1mm}         \\
\hline
variant A      &    7         & 	2.5    &   0.17       &     1.17                 \\
variant B      &    7         & 	1.5    &   0.11       &     1.54                 \\
variant C      &    7         & 	8.0      &   0.15       &     1.00                 \\
variant D      &    7         & 	11.5   &   0.13       &     1.26                 \\
 \end{tabular}
 \end{ruledtabular}
 \label{tab:parameters}
\end{table}

\section{Conclusions}
\label{sec:summary}

The structural, electronic, and dynamical properties of the Fe$_3$Si/GaAs(001) heterostructure were investigated within the density functional theory.
Four different atomic configurations of the Fe$_3$Si/GaAs interface were studied and their stability was analyzed by calculating the formation energies and phonon dispersion curves.
The most stable are variants B and C with pure Fe and mixed Fe-Si interface layers, respectively.
Variant B shows full dynamical stability. Configuration D is dynamically unstable and has the lowest binding energy.
However, we cannot rule out the stabilization of any variant at finite temperature, as the DFT calculations correspond to T = 0~K. 
Relatively small differences between the formation energies 
of the studied systems also indicate that all variants may be observed.

We studied the influence of the interface on magnetization and charge distribution. We found a transfer of electrons from the first Fe-Si or Fe layer into the substrate GaAs layer, reducing electron localization in the interface areas.
We have used two different methods for calculating magnetic moments and obtained consistently very similar values for the Fe atoms.
In the first layer from the interface, they are enhanced in comparison with bulk Fe$_3$Si, and  the strongest effect is found in the pure Fe layers in variants B and D.

Calculations with magnetic moments fixed along different axes showed that the energetically favorable direction of magnetization is the [110] direction of the Fe$_3$Si crystallographic unit cell, in agreement with experimental results. Moreover, we demonstrate that both a tetragonal distortion and an interaction with the substrate are responsible for the magnetic moments alignment.  

The electronic densities of states projected on Fe and Si atoms belonging to different layers demonstrate a strong modification of spectra calculated for atoms in close vicinity to the substrate. 
The changes observed for spin-down states at the Fermi energy affect the spin polarization.  
In configurations A and D, the spin polarization at the Fermi energy is larger than the bulk value for the Fe$_{3}$Si crystal. 
This supports the suitability of the Fe$_3$Si/GaAs interface as a possible building block in heterostructures for future magnetoelectronic and spintronic applications.

The effect of the GaAs substrate on the lattice vibrations of the Fe$_3$Si layers is also discussed. In all interface configurations, we observe a pronounced shift to lower energies of the PDOS of the atoms located at the interface. This shift is especially strong in the case of the in-plane vibrations. In further layers, vibrational spectra are similar to the PDOS of the corresponding layers in the bulk.

Phonon spectra obtained for different configurations of the Fe$_{3}$Si/GaAs interface were used to analyze the Fe-partial PDOS measured for the Fe$_3$Si/GaAs heterostructure with a 3 monolayers thick Fe$_3$Si layer.
Due to the grazing-incidence scattering geometry of the experiment, we considered only the in-plane Fe vibrations. 
In the measured spectrum, the sharp peaks present in the theoretically obtained PDOS are strongly broadened, however, both spectra cover the same energy range. The broadening of peaks may be caused by crystal defects at the epitaxial Fe$_{3}$Si/GaAs interface.
We analyzed the correspondence of the experimental and theoretical spectra comparing  the contributions from different variants.
The best agreement was found for variant C, which reproduces the low-energy states observed experimentally below 10 meV.
The anomalous enhancement of the phonon density of states at low energies is explained by the interface-specific phonon modes.

\begin{acknowledgments}

The authors are thankful to Konrad J. Kapcia for very fruitful discussions and comments.
This work was supported by the National Science Centre (NCN, Poland) under Grant No. UMO-2017/25/B/ST3/02586.
S.S. acknowledges financial support by the Helmholtz Association (VH-NG-625) and BMBF (05K16VK4).
\end{acknowledgments}

\bibliography{refs}

\end{document}